\documentclass{aa}
\usepackage{epsfig}
\usepackage{amsmath}
\usepackage{txfonts}
\usepackage{natbib}
\usepackage{bm}

\begin{document}

\title{Excitation and evolution of coronal oscillations in self-consistent 3D radiative MHD simulations of the solar atmosphere}
\authorrunning{Kohutova \& Popovas}
\titlerunning{Excitation of coronal oscillations in 3D MHD simulations}
\author{P. Kohutova\inst{1,2} \and A. Popovas\inst{1,2}}
\institute{Rosseland Centre for Solar Physics, University of Oslo, P.O. Box 1029, Blindern, NO-0315 Oslo, Norway\\
\email{petra.kohutova@astro.uio.no}
\and
Institute of Theoretical Astrophysics, University of Oslo, P.O. Box 1029, Blindern, NO-0315 Oslo, Norway}
\date{Received; accepted}

\abstract
{Solar coronal loops are commonly subject to oscillations. Observations of coronal oscillations are used to infer physical properties of the coronal plasma using coronal seismology.} 
{Excitation and evolution of oscillations in coronal loops is typically studied using highly idealised models of magnetic flux-tubes. In order to improve our understanding of coronal oscillations, it is necessary to consider the effect of realistic magnetic field topology and evolution.}
{We study excitation and evolution of coronal oscillations in three-dimensional self-consistent simulations of solar atmosphere spanning from convection zone to solar corona using radiation-MHD code Bifrost. We use forward-modelled EUV emission and three-dimensional tracing of magnetic field to analyse oscillatory behaviour of individual magnetic loops. We further analyse the evolution of individual plasma velocity components along the loops using wavelet power spectra to capture changes in the oscillation periods.}
{Various types of oscillations commonly observed in the corona are present in the simulation. We detect standing oscillations in both transverse and longitudinal velocity components, including higher order oscillation harmonics. We also show that self-consistent simulations reproduce existence of two distinct regimes of transverse coronal oscillations: rapidly decaying oscillations triggered by impulsive events and sustained small-scale oscillations showing no observable damping. No harmonic drivers are detected at the footpoints of oscillating loops.}
{Coronal loop oscillations are abundant in self-consistent 3D MHD simulations of the solar atmosphere. The dynamic evolution and variability of individual magnetic loops suggest we need to reevaluate our models of monolithic and static coronal loops with constant lengths in favour of more realistic models.}


\maketitle

\section{Introduction}
Coronal loops, the basic building blocks of solar corona, are commonly subject to oscillatory behaviour. Due to their magnetic structure they act as wave guides and can support a variety of wave modes \citep[see e.g. reviews by][]{nakariakov_2005,nakariakov_2020}. Various types of wave modes have been reported at coronal heights; these include nearly incompressible transverse oscillations in the direction perpendicular to the loop axis \citep{aschwanden_1999, nakariakov_1999, verwichte_2013}, compressible longitudinal oscillations along the loop axis \citep{berghmans_1999, moortel_2000} and incompressible torsional oscillations \citep{kohutova_2020a}. Transverse oscillations of coronal loops are by far the most commonly observed type. They were first discovered in the TRACE observations of coronal loops following a solar flare \citep{aschwanden_1999, nakariakov_1999} and since then have been studied extensively through high-resolution solar observations, numerical simulations and analytical theory. The most commonly reported oscillation periods lie in the range of 1 to 10 min, and the oscillation amplitudes range from few hundred km to several Mm. Most of the observed transverse coronal loop oscillations have been identified as corresponding to either standing or propagating kink modes. This mode identification is based on modelling individual coronal loops as cylindrical flux tubes \citep{edwin_1983}. It should however be noted that the density structuring in the corona is more complex than this and assumptions about the individual mode properties based on idealised modelling of homogeneous plasma structures do not always apply in a non-homogeneous and dynamic environment that is real solar corona \citep[e.g.][]{goossens_2019}. The fundamental harmonic of standing kink mode is the most commonly observed, where the footpoints of the coronal loop act as nodes and the point of maximum displacement amplitude lies at the loop apex. Despite fundamental harmonic being the most intuitive resonant response of a line-tied magnetic fieldline to an external perturbation, excitation of higher-order harmonics is also perfectly viable; higher order harmonics have also been identified in coronal loop oscillations \citep{verwichte_2004, duckenfield_2018}.

Transverse oscillations of coronal loops are commonly used for coronal seismology \citep[see e.g. reviews by][]{moortel_2005, de_moortel_2012, nakariakov_2020}. Coronal seismology is a powerful method that relies on using oscillation parameters that are readily observable in the coronal imaging and spectral data to deduce plasma properties such as density, temperature, magnetic field strength or density structuring, which are otherwise very challenging to measure directly. The accuracy of the seismologically deduced quantities is however limited by somewhat simplifying assumptions of modelling coronal structures as homogeneous plasma cylinders and is subject to uncertainties caused by projection and line-of-sight effects.

Transverse coronal loop oscillations are often classified as occurring in two regimes: large-amplitude rapidly damped oscillations and small amplitude seemingly 'decayless' oscillations. Large amplitude oscillations are typically triggered by a external impulsive event, which is usually easy to identify, such as blast wave following a solar fare \citep[e.g.][]{white_2012}. Amplitudes are of the order of 1 Mm and the oscillations are rapidly damped within 3-4 oscillation periods. 

Small amplitude oscillations are observed without any clearly associated driver and can persist for several hours \citep{wang_2012, nistico_2013, anfinogentov_2015}. Their amplitudes are of the order of 100 km and they show no observable damping. Even though it is clear they must be driven by some small-scale abundant process, the exact excitation mechanism of this oscillation regime is unclear. Several excitation mechanisms have been proposed to explain the persistent nature of these oscillations. These include mechanisms acting at coronal heights such as onset of thermal instability in the corona \citep{kohutova_2017, verwichte_2017}, presence of siphon flows in the coronal loop \citep{kohutova_2018}, self-oscillation of coronal loops due to interaction with quasi-steady flows \citep{nakariakov_2016, karampelas_2020} as well as footpoint-concentrated drivers. Commonly assumed excitation mechanism is associated with turbulent vortex flows which are ubiquitous in the lower solar atmosphere \citep{carlsson_2010, shelyag_2011, liu_2019} and the resulting random footpoint buffeting. This is typically modelled as a stochastic driver at the footpoints \citep[e.g.][]{pagano_2019}. Such driver however does not reproduce the observational characteristics of this type of oscillations, in particular the stability of the oscillation phase and lack of any observable damping \citep{nakariakov_2016}. It has been proposed that a footpoint driver of the form of a broadband noise with a power-law fall-off can in principle lead to excitation of eigenmodes along coronal loops \citep{afanasyev_2020}, this has however not been tested through MHD simulations. Finally, excitation of transverse oscillations through solar p-mode coupling has also been proposed in several studies \citep{tomczyk_2009, morton_2016, morton_2019, riedl_2019}. Since p-modes correspond to acoustic waves generated by turbulent flows in the solar convection zone, they are subject to acoustic cut-off at the $\beta = 1$ equipartition layer. High magnetic field inclinations can however lead to p-mode leakage into chromosphere even for frequencies below the cut-off frequency \citep{pontieu_2004}, and to subsequent coupling to different magnetoacoustic wave modes \citep{santamaria_2015}. A 3D MHD simulations of coronal fluxtubes with gravity-acoustic wave driver at the footpoints however did not find any evidence of the acoustic driving leading to significant transverse displacement of the fluxtubes at coronal heights \citep{riedl_2019}. 

It is in fact very likely that the oscillatory behaviour observed in the corona is caused by a combination of several different excitation mechanisms. 3D MHD simulations of solar atmosphere spanning from convection zone to corona in principle include all of the potential mechanisms discussed above. However, so far only impulsively generated rapidly damped coronal loop oscillations have been studied in such 3D MHD simulations \citep{chen_2015}, which show good match of damping rates compared with observed impulsively generated oscillations. Such numerical works also highlight the implications of observational limitations for coronal seismology, as they enable direct comparison of seismologically deduced and real values of physical quantities. Detailed properties of transverse coronal loop oscillations have been studied numerically in great detail in recent years, including their evolution, damping and development of instabilities and resulting nonlinear dynamics \citep[e.g.][]{antolin_2014, magyar_2016, karampelas_2017}. Most of these studies however employ simplified geometries that model coronal structures as straight uniform fluxtubes. Additionally, such studies also rely on artificially imposed harmonic footpoint drivers \citep{karampelas_2017, pagano_2017}, without discussing where do these drivers originate from. Such approach makes it possible to isolate individual effects of interests, and focus on evolution at small scales through techniques such as adaptive mesh refinement \citep[e.g.][]{magyar_2015} due to obvious computational limitations that come with studying physical mechanisms operating at vastly different spatial scales. There is, however, need for more complex approach to see how such effects manifest in a realistic solar atmosphere setup and to identify their observational signatures.

In this work we investigate the presence of oscillations of coronal loops in a simulation spanning from convection zone to corona, with a realistic lower solar atmosphere dynamics, magnetic field geometry and density structuring. We for the first time analyse excitation of transverse coronal oscillations in a self-consistent simulation of the solar atmosphere. We use combination of forward modelling of coronal EUV emission and evolution of physical quantities along magnetic field to determine oscillation characteristics.

\begin{figure*}
	\includegraphics[width=43pc]{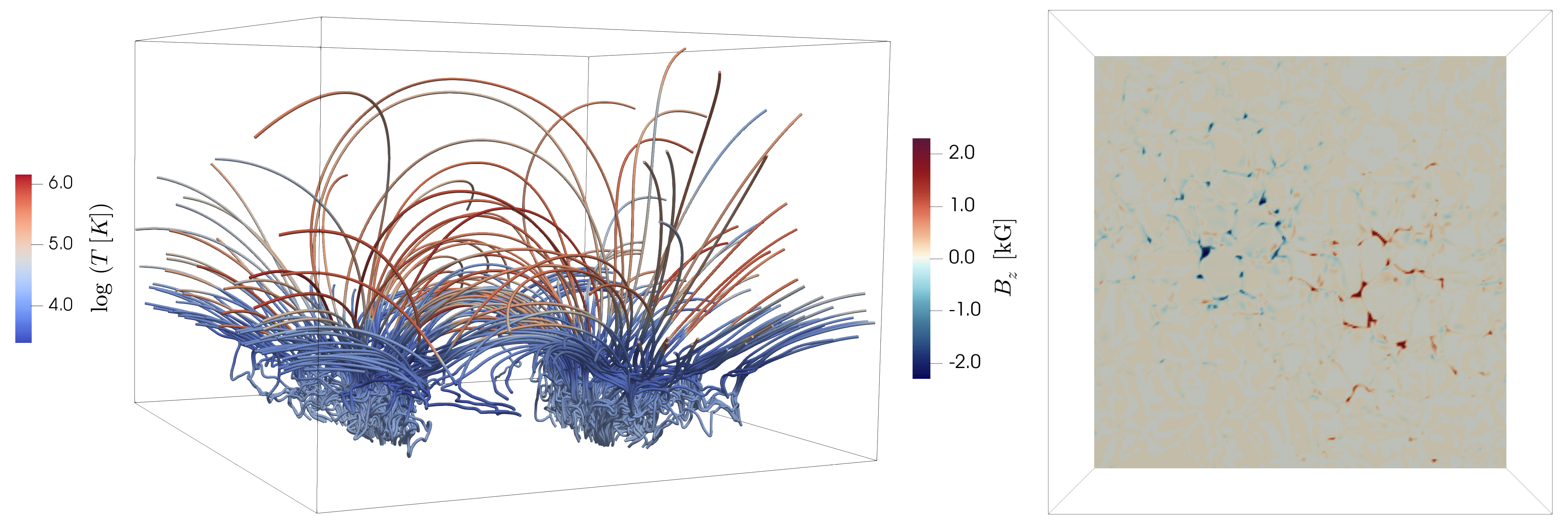}
	\caption{Left: Magnetic configuration of the simulation showing the temperature profile of the individual fieldlines. The simulation domain has physical size of 24 $\times$ 24 $\times$ 16.8 Mm. The simulation snapshot shown corresponds to $t = 180$ s after the non-equilibrium ionisation of hydrogen has been switched on. Right: Line-of-sight component $B_{\mathrm{z}}$ of the photospheric magnetic field at $t = 180$ s.}
	\label{fig:context_field}
\end{figure*}

\begin{figure*}
	\includegraphics[width=43pc]{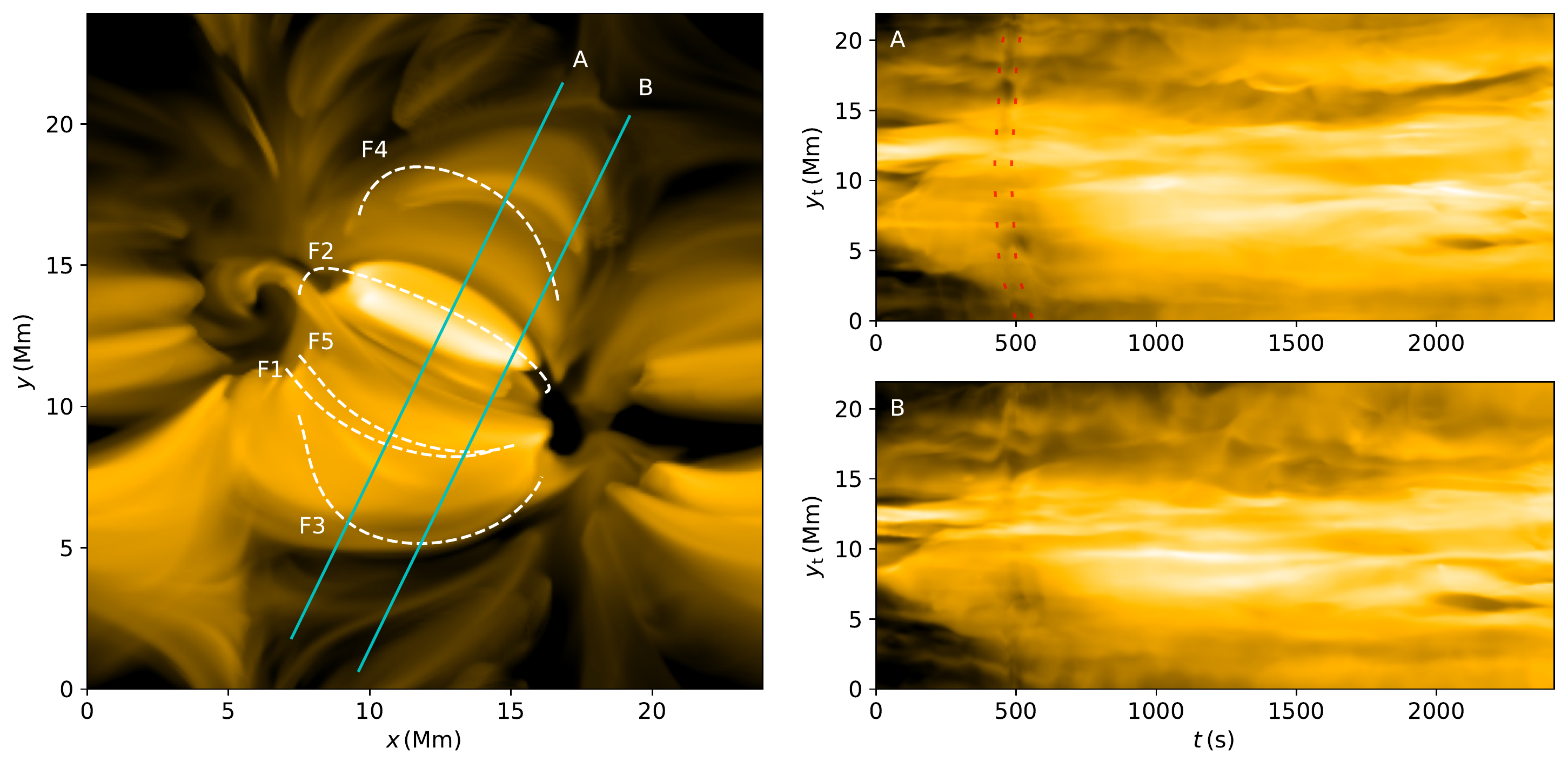}
	\caption{Left: forward-modelled emission in the SDO/AIA 171 {\AA} channel at at $t = 180$ s. Dashed lines outline projected positions of the studied fieldlines F1 - F5. Right: Time-distance plots showing temporal evolution along the slits shown in blue. The vertical axis corresponds to the distance along the slit. Several oscillating structures are visible in the time-distance plots. Dashed contour in plot A outlines a propagating disturbance.} Animation of this figure is available.
	\label{fig:fomo}
\end{figure*}

\begin{figure}
	\includegraphics[width=21pc]{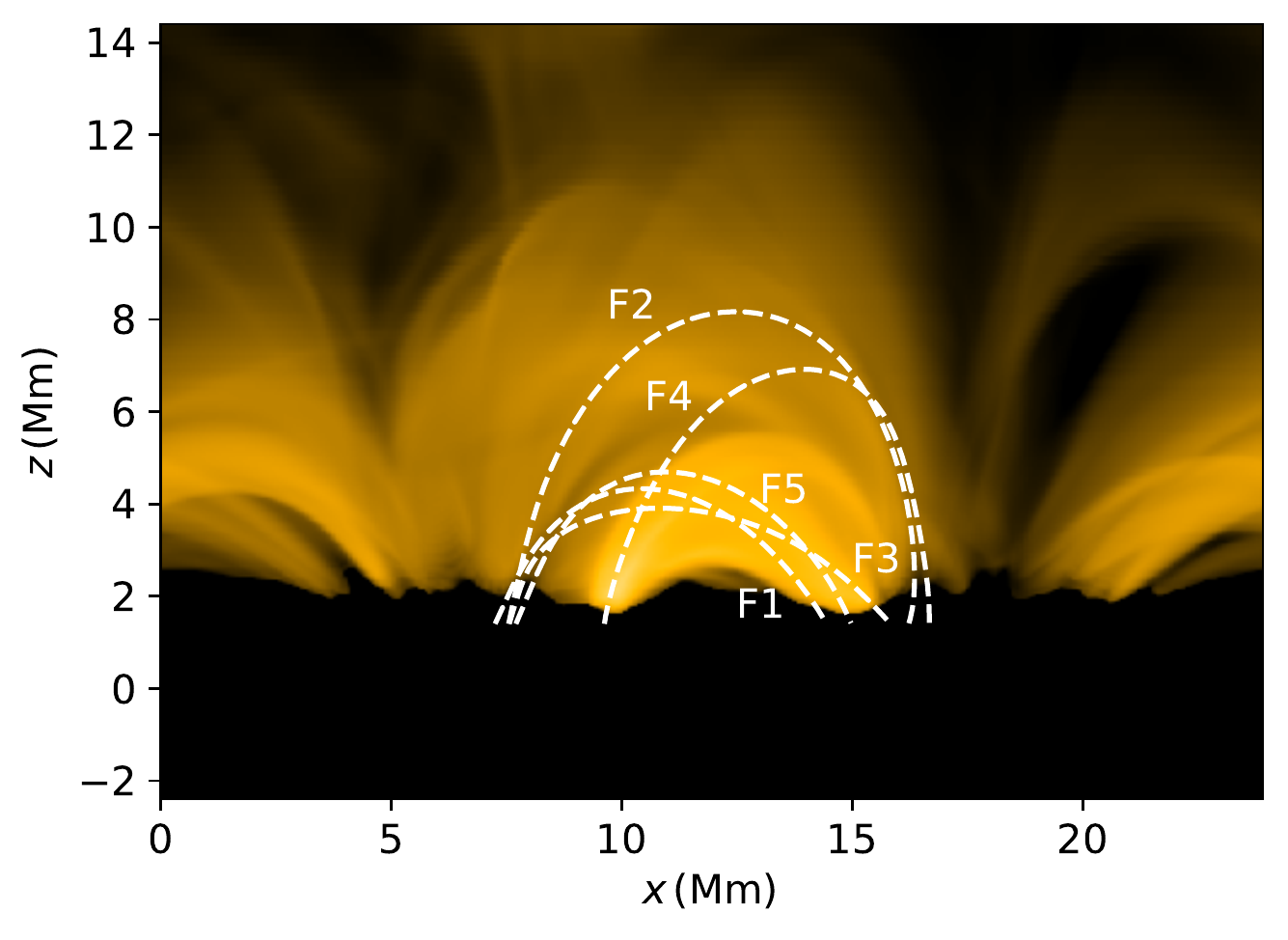}
	\caption{Forward-modelled emission in the SDO/AIA 171 {\AA} channel at at $t = 180$ s viewed along y-axis. Dashed lines outline projected positions of the studied fieldlines F1 - F5.}
	\label{fig:fomo_side}
\end{figure}

\section{Numerical Model}
\label{section:model}
We analyse coronal oscillations in the numerical simulation of magnetic enhanced network spanning from the upper convection zone to the corona \citep{carlsson_2016} using 3D radiation MHD code Bifrost \citep{gudiksen_2011}. This simulation has been previously used for analysing oscillatory behaviour of chromospheric fibrils by \citet{leenaarts_2015} and for studying signatures of propagating shock waves by \citet{eklund_2020}. 

Bifrost solves resistive MHD equations and includes non-LTE radiative transfer in the photosphere and low chromosphere and parametrized radiative losses and heating in the upper chromosphere, transition region and corona. The simulation includes the effects of thermal conduction parallel to the magnetic field and the non-equilibrium ionization of hydrogen.

The physical size of the simulation box is 24 $\times$ 24 $\times$ 16.8 Mm. The vertical extent of the grid spans from 2.4 Mm below the photosphere to 14.4 Mm above the photosphere, which corresponds to $z=0$ surface and is defined as the (approximate) height where the optical depth $\tau_{500}$ is equal to unity. The simulation is carried out on a 504 $\times$ 504 $\times$ 496 grid. The grid is uniform in the $x$ and $y$ direction with grid spacing of 48 km. The grid spacing in $z$ direction is non-uniform in order to resolve steep gradients in density and temperature in the lower solar atmosphere and varies from 19 km to 98 km. The domain boundaries are periodic in the $x$ and $y$ direction and open in the $z$ direction. The top boundary uses characteristic boundary conditions such that any disturbances are transmitted through the boundary with minimal reflection. Tests of the characteristic boundaries implemented in Bifrost suggest that the reflection of the energy from the upper boundary is 5\% or less \citep{gudiksen_2011}. At the bottom boundary the flows are let through and the magnetic field is passively advected while keeping the magnetic flux through the bottom boundary constant. The average horizontal pressure is driven towards a constant value with a characteristic timescale of 100 s. This pressure node at the bottom boundary gives rise to acoustic wave reflection which resembles the refraction of waves in the deeper solar atmosphere. This leads to global radial box oscillations with a period of 450 s, which are a simulation counterpart of Solar p-modes \citep{stein_2001, carlsson_2016} (radial modes in this context correspond to oscillations in horizontally averaged quantities polarised along z-axis in a cartesian simulation box).

The photospheric magnetic field has average unsigned value of about 50 G and is concentrated in two patches of opposite polarity about 8 Mm apart; this leads to development of several magnetic loops at coronal heights (Fig. \ref{fig:context_field}). Upon initialisation of the simulation, the large-scale magnetic field configuration was determined by the use of potential extrapolation of the vertical magnetic field, specified at the bottom boundary. After inserting the magnetic field into the domain it is quickly swept around by convective motions. These lead to magnetic field braiding and associated Ohmic and viscous heating which together maintain high temperature in the chromosphere and corona (see e.g. \citet{carlsson_2016, kohutova_2020b} for further details of the numerical setup).

\begin{figure*}
	\includegraphics[width=43pc]{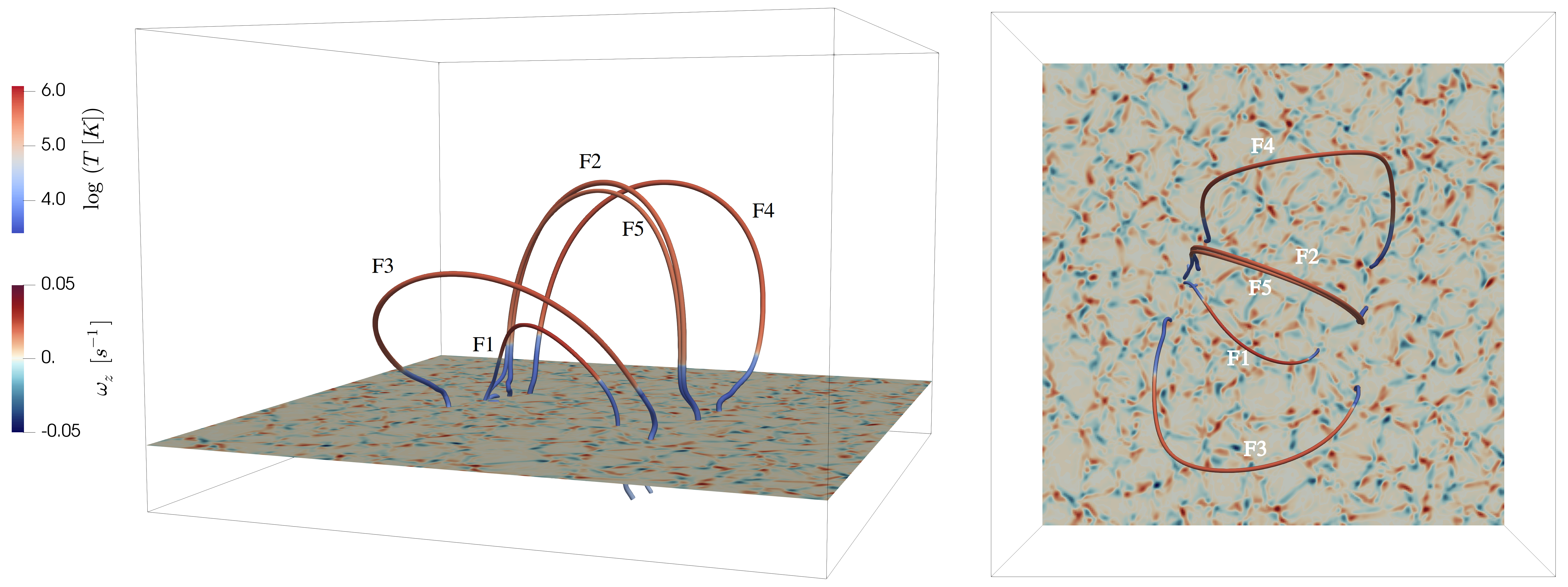}
	\caption{Left: Location of fieldlines F1 - F5 chosen for analysis of oscillatory behaviour shown at $t = 1500$ s. We also show z-component of vorticity $\omega_{z}$ at z = 0, i.e. in the photosphere. The footpoints of analysed fieldlines are embedded in the intergranular lanes with strong vortex flows. Right: Configuration viewed from above. Animation of this figure is available.}
	\label{fig:context_loops}
\end{figure*}

\section{Transverse oscillations observable in forward-modelled EUV emission}

In order to compare oscillatory behaviour seen in the numerical simulation with the commonly observed transverse coronal loop oscillations seen by SDO/AIA, we forward model the EUV emission using the FoMo code \citep{doorsselaere_2016}. The FoMo code is capable of modelling optically thin coronal emission in individual spectral lines as well as in the broadband SDO/AIA channels. The instrumental response function of a given SDO/AIA bandpass $\kappa_{\alpha}$ is calculated for a grid of electron densities $n_{\mathrm{e}}$ and temperatures $T$ as

\begin{equation}
\kappa_{\alpha}(n_{\mathrm{e}}, T) = \int G(\lambda_{\alpha}, n_{\mathrm{e}}, T) R_{\alpha}(\lambda_{\alpha}) \mathrm{d}\lambda_{\alpha}
\end{equation}

where $G(\lambda_{\alpha}, n_{\mathrm{e}}, T)$ is the contribution function calculated from the CHIANTI database, $R_{\alpha}(\lambda_{\alpha})$  is the wavelength-dependent response function of the given bandpass, and the wavelength integration is done over all spectral lines within the given bandpass. $\kappa_{\alpha}$ is then integrated along the line-of-sight parallel to z-axis, which corresponds to looking straight down on the solar surface. 

Figure \ref{fig:fomo} shows forward modelled emission in the SDO/AIA 171 {\AA} bandpass, which is dominated by the Fe IX line with formation temperature $\log T = 5.9$. In order to highlight the oscillatory motion, a time-distance plot is created by taking the intensities along a slit across the loop structures and stacking them in time. Several oscillating structures can be seen in such plot, suggesting transverse coronal loop oscillations are abundant in the model. The forward-modelled EUV emission in the model is more diffuse (Fig. \ref{fig:fomo_side}) and subject to less observable structuring across the magnetic field than coronal emission typically observed at similar wavelengths, where the individual coronal loops can appear very threadlike \citep[e.g.][]{peter_2013, williams_2020}. We note that the simulation does not include magnetic flux emergence which means the loops are not mass-loaded and pushed into the corona from the lower solar atmosphere. Instead, the dense loops are filled via chromospheric evaporation caused by localised heating \citep{kohutova_2020b}. Some transverse structuring is however still present thanks to the temperature variation across different magnetic fieldlines. This makes it possible to see oscillatory behaviour of the individual coronal strands in the intensity time-distance plot.

\section{Transverse oscillations of individual field lines}

\begin{table}[t]
	\caption{Loop properties}
	\label{tab:speeds}
	\centering
	\tabcolsep=0.11cm
	\begin{tabular}{ccccccccc}
		\hline\hline
		FL & $L$ & $\bar{\rho}$ & $\bar{T}$ &  $\bar{B}$ & $C_S$ & $V_A$ \\ 
		& (Mm) &  (kg m$^{-3}$) & (K) & (G) &  (km s$^{-1}$) &  (km s$^{-1}$)\\
		\hline
		$\mathrm{F1}$  & 10.1 & $5.6 \times 10^{-12}$ & $7.2 \times 10^{5}$ & 26 & 129 & 980   \\
		$\mathrm{F2}$  & 17.0 & $2.2 \times 10^{-12}$ & $5.7 \times 10^{5}$ & 17 & 114 & 989   \\
		$\mathrm{F3}$  & 15.2 & $1.5 \times 10^{-12}$ & $5.5 \times 10^{5}$ & 9 & 113 & 687   \\
		$\mathrm{F4}$  & 17.3 & $1.2 \times 10^{-12}$ & $4.3 \times 10^{5}$ & 11 & 99 & 910  \\
		$\mathrm{F5}$  & 12.8 & $2.8 \times 10^{-12}$ & $6.6 \times 10^{5}$ & 22 & 123 & 1186  \\
		
		\hline
	\end{tabular}
	\tablefoot{Average physical quantities along individual fieldlines, and the corresponding sound speeds ($C_S$) and  Alfv\'{e}n speeds ($V_A$).}
\end{table}

\begin{figure*}
	\includegraphics[width=43pc]{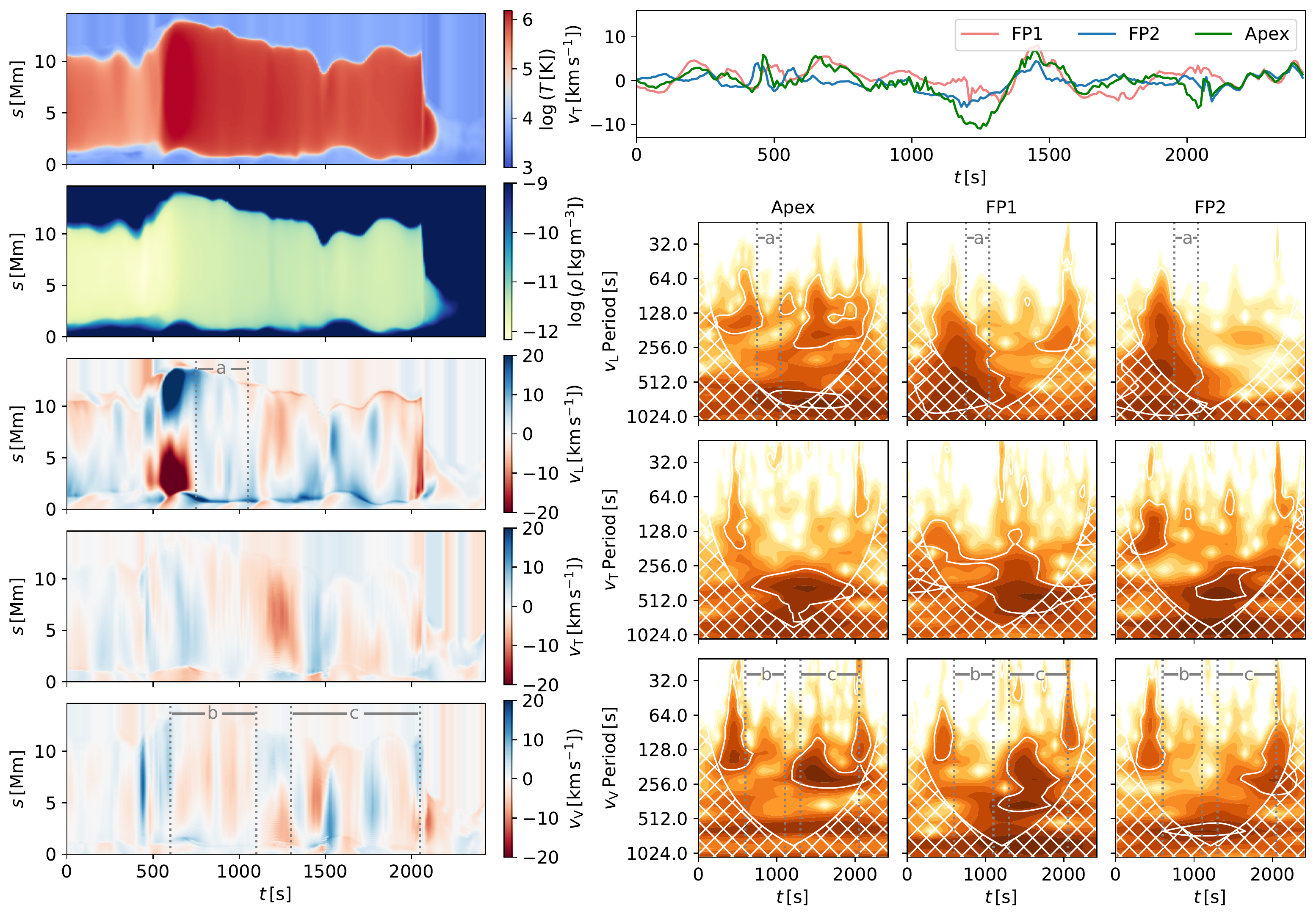}
	\caption{Left: Evolution of temperature, density and three velocity components $v_{\mathrm{L}}$, $v_{\mathrm{T}}$ and $v_{\mathrm{V}}$ along the fieldline F1. The $x$ axis corresponds to time and the $y$ axis corresponds to the position along the fieldline. Vertical dotted lines mark segments with oscillatory behaviour. Top right: Evolution of the transverse component of the velocity $v_{\mathrm{T}}$ at the footpoints FP1 and FP2 (red and blue respectively) 1 Mm above the transition region and at the fieldline apex halfway between the two footpoints (green). Bottom right: Wavelet spectra for the 3 velocity components taken at the apex of the fieldline and at the footpoints FP1 and FP2. Dark colour corresponds to high power. The hatched region corresponds to the area outside of the cone of influence. White lines correspond to 95 \% significance contours.}
	\label{fig:f1}
\end{figure*}

\begin{figure*}
	\includegraphics[width=43pc]{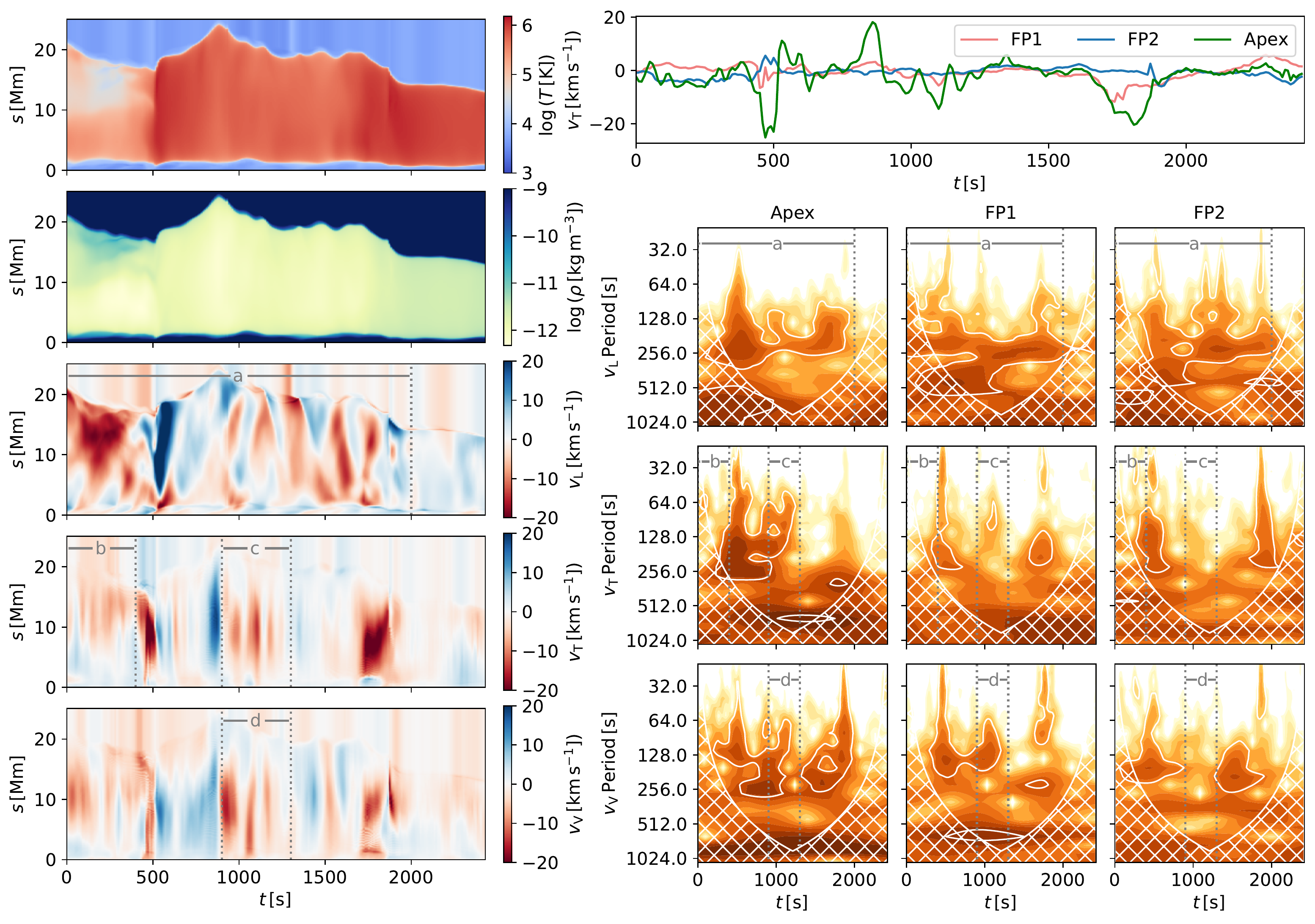}
	\caption{Same as Fig. \ref{fig:f1}, but for fieldline F2.}
	\label{fig:f2}
\end{figure*}

\begin{figure*}
	\includegraphics[width=43pc]{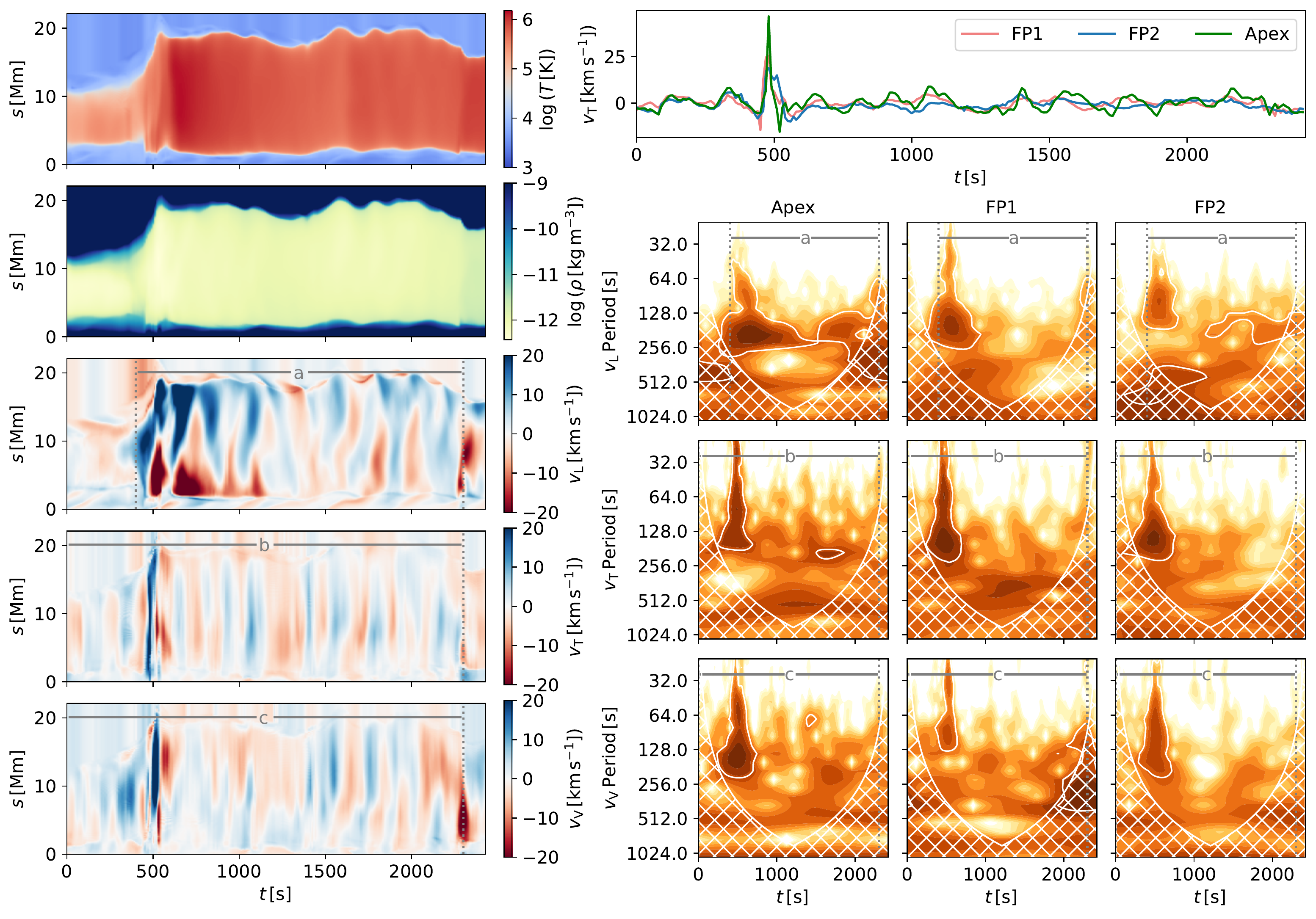}
	\caption{Same as Fig. \ref{fig:f1}, but for fieldline F3.}
	\label{fig:f3}
\end{figure*}

\begin{figure*}
	\includegraphics[width=43pc]{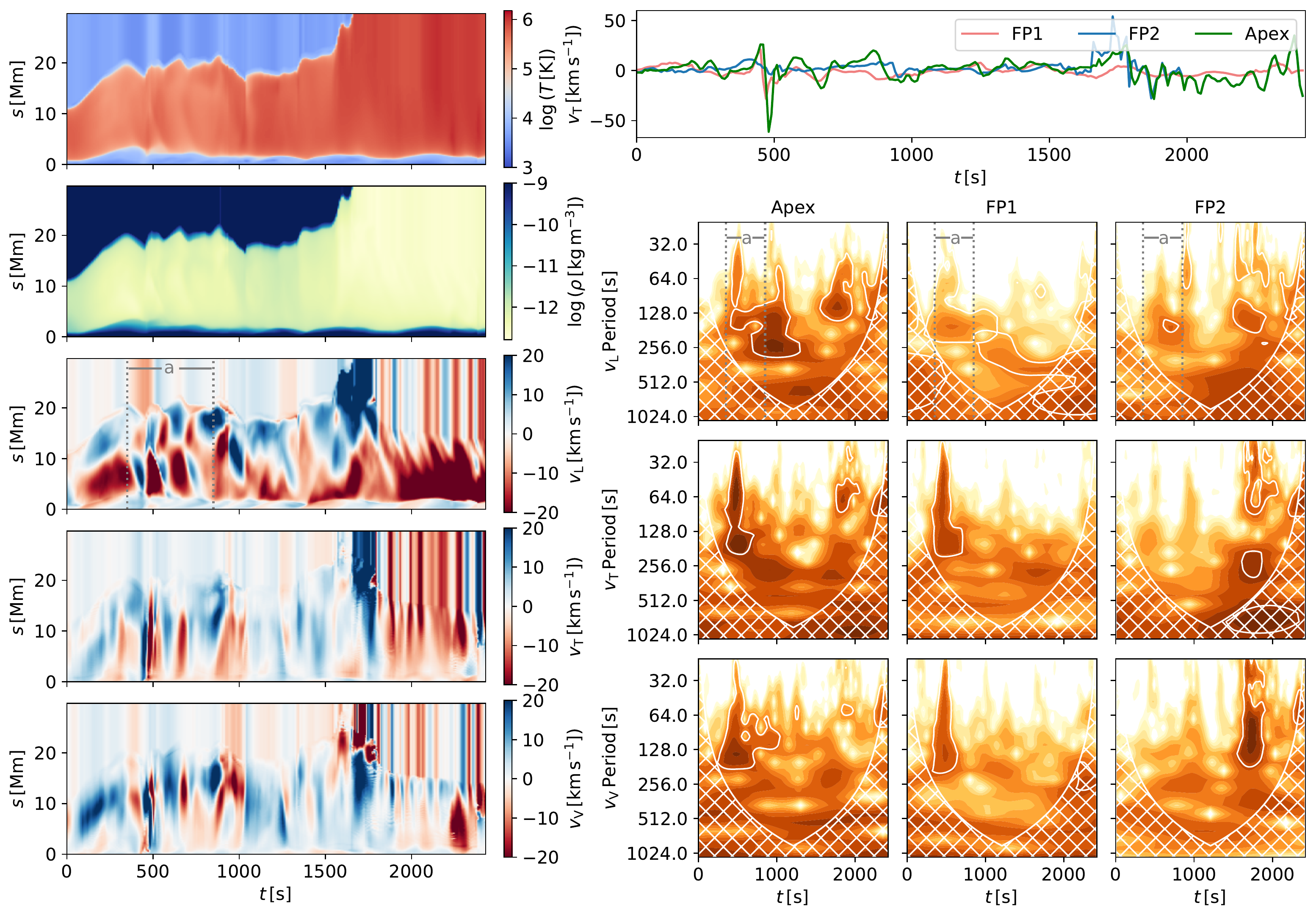}
	\caption{Same as Fig. \ref{fig:f1}, but for fieldline F4.}
	\label{fig:f4}
\end{figure*}

\begin{figure*}
	\includegraphics[width=43pc]{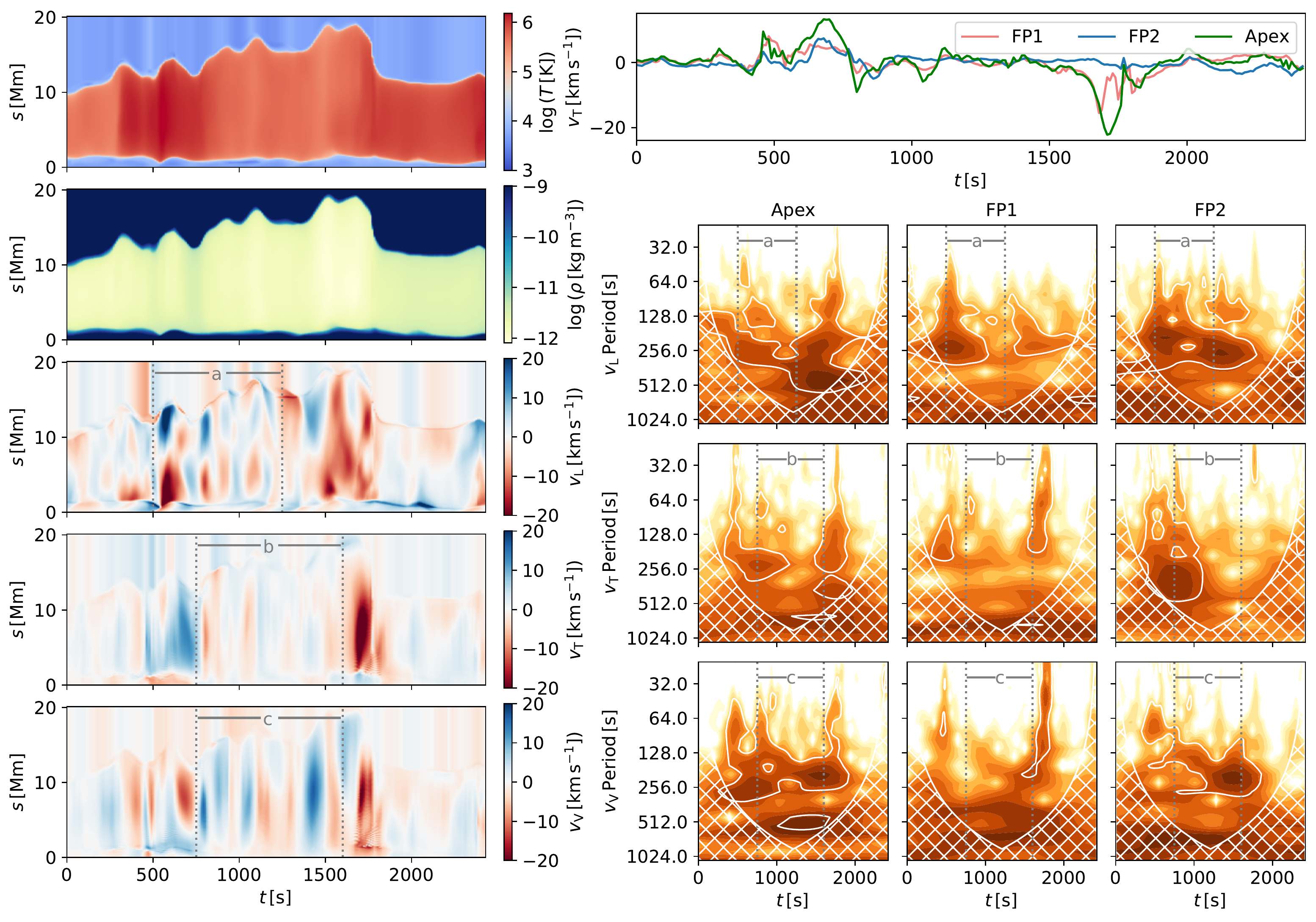}
	\caption{Same as Fig. \ref{fig:f1}, but for fieldline F5.}
	\label{fig:f5}
\end{figure*}

Coronal loops in our study are represented by closed magnetic fieldlines extending into the corona. Several of those have enhanced densities compared to the surroundings. Density-enhanced coronal loops with well-defined boundaries in the simulation are however less common than what is usually observed in the corona. As there is no flux emergence mechanism included in the simulation, most of the dense loops are filled by chromospheric evaporation instead of being lifted up from the lower atmosphere. 

The magnetic field in the simulation domain is constantly evolving and undergoing complex motions which include sideways displacement, oscillatory motion, torsional and upward/downward motion resulting in the change of the total coronal loop length. Therefore, in order to investigate the oscillatory behaviour of coronal loops in the simulation, it is necessary to trace the evolution of the corresponding magnetic field lines through both time and space. To do this, we use a field tracing method previously used by \citet{leenaarts_2015} and \citet{kohutova_2020b}. A magnetic fieldline is defined as a curve in 3D space $\vec{r}(s)$ parametrised by the arc length along the curve $s$, for which $\mathrm{d} \vec{r}/ \mathrm{d} s = \vec{B} / |\vec{B}|$.

The tracing of the evolution of magnetic fieldlines is done by inserting seed points (one seedpoint per tracked fieldline) into the locations in the simulation domain which show oscillatory behaviour in the velocity time-distance plots. This is equivalent to tracking the evolution of magnetic conectivity of an oscillating plasma element. The seed points are then passively advected forward and backward in time using the velocity at the seed point position. At every time step the magnetic field is then traced through the instantaneous seed point position in order to determine the spatial coordinates of the traced fieldline. This is done using the Runge–Kutta–Fehlberg integrator with an adaptive step size. Even though the accuracy of this method is limited by the size of the time step between the two successive snapshots (i.e. 10 seconds), it works reasonably well provided that the magnetic field evolution is smooth and there are no large amplitude velocity variations occurring on timescales shorter than the size of the time step. We note that this method leads to a jump in the fieldline evolution in the instances where magnetic reconnection occurs.

We investigate the evolution of 5 different fieldlines labelled F1...F5 (Fig. \ref{fig:context_loops}). We note that the footpoints of chosen magnetic loops lie in the regions of enhanced vorticity. We analyse the evolution of the temperature, density, and 3 velocity components, $v_{\mathrm{L}}$, $v_{\mathrm{V}}$ and $v_{\mathrm{T}}$ along each loop. The longitudinal velocity $v_{\mathrm{L}} = \vec{v} \cdot \vec{T}$ corresponds to the velocity component aligned with the tangent vector of the magnetic fieldline given by $\vec{T} = \vec{B}/|\vec{B}|$. The vertical velocity $v_{\mathrm{V}} = \vec{v} \cdot \vec{N}$ is the velocity component along the fieldline normal vector given by $\vec{N} =  \frac{\mathrm{d}\vec{T}}{\mathrm{d}s} / |\frac{\mathrm{d}\vec{T}}{\mathrm{d}s}| $ and corresponds to the motion in the plane of the coronal loop. Finally the transverse velocity component along the binormal vector is given by $v_{\mathrm{T}} =  \vec{v} \cdot \vec{R} $ where $\vec{R} = \vec{T} \times \vec{N}$ and corresponds to transverse motion perpendicular to the plane of the coronal loop. Unit vectors $\vec{T}$, $\vec{N}$ and $\vec{R}$ together form an orthogonal coordinate system equivalent to a Frenet frame of reference. Such coordinate frame is well suited for analysing oscillations in complex 3D magnetic field geometries and is commonly used in such studies \citep[e.g.][]{felipe_2012, leenaarts_2015, gonzalez_2019}. We further calculate a wavelet power spectrum \citep{torrence_1998} for all three velocity components at 3 different locations along the fieldline; close to the loop footpoints 1 Mm above the transition region (height of which is tracked with time) at the beginning and end of the fieldline (labelled FP1 and FP2 for left and right footpoint respectively) and at the loop apex halfway between FP1 and FP2.

\subsection{Fieldline F1}
Fieldline F1 is a short, closed fieldline in the centre of the domain. For the most of its lifetime it is not subject to any major changes, however at $t = 2100$ s it rapidly contracts and finally disappears in the chromosphere. The evolution of the physical quantities along the fieldline is shown in Fig. \ref{fig:f1}. The fieldline is subject to strong base heating starting at t $\sim$ 600 s resulting in evaporative upflows of the chromospheric plasma into the loop, observable in the $v_{\mathrm{L}}$ component for about 200 s. We identify this as a consequence of a global disturbance propagating from the centre of the domain outwards, visible in Fig. \ref{fig:fomo}), which likely triggers reconnection in the affected loops, leading to increased Joule and viscous heating. This is accompanied by an onset of the oscillatory behaviour lasting from $t = 600$ s to $t = 1100$ s. The oscillation is most clearly seen in the $v_{\mathrm{V}}$ component of the velocity (marked as segment \textit{b} in Fig. \ref{fig:f1}) which corresponds to vertical polarisation (i.e. polarisation in the plane of the loop). In total, 3 oscillation periods are observable, with the oscillation period being around 200 s. The lack of periodic displacement at the loop footpoints as compared with the periodic evolution at the loop apex lacking phase shift indicate that the oscillation is standing, and not driven by footpoint motions. This is accompanied by an oscillation with matching period in $v_{\mathrm{L}}$ velocity component, that is along the magnetic field, with the oscillation profile matching the second harmonic of a standing longitudinal mode (segment \textit{a} in Fig. \ref{fig:f1}). The wavelet analysis also shows presence of an oscillation with 250 s period in the $v_{\mathrm{V}}$ with later onset having similar characteristics and lasting from $t=1300$ s to $t = 2100$ s, that is until the impulsive event that leads to rapid contraction of the loop. This oscillation is marked as segment \textit{c} in the $v_{\mathrm{V}}$ evolution plot. Some attenuation of the oscillation is observable, especially in the initial stages. We attribute the increase of the standing oscillation period to increase in density of the coronal loop plasma. Finally, a large amplitude disturbance ($\sim$ 10 km s$^{-1}$ amplitude at the apex) can be observed in the evolution of the transverse velocity component $v_{\mathrm{T}}$ from $t=1200$ s to $t = 1600$ s. However, as only one oscillation period is observable, we refrain from drawing any conclusions about the oscillation characteristics.

\subsection{Fieldline F2}
Fieldline F2 has length of over 20 Mm (Fig. \ref{fig:f2}). During the initial 400 s of its evolution, there is a clear accumulation of cool and dense material in the coronal part of the loop, most likely through the condensation of coronal plasma. Condensation phenomena in this simulation has been studied in detail by \citet{kohutova_2020b}. The loop evolution is dynamic and affected by impulsive events occurring at coronal heights. At $t\sim 500$ s a discontinuity in temperature and velocity evolution is visible similar to what is observed in F1. Associated temperature increase and jump in fieldline tracking suggest this corresponds to a reconnection event. We note that the discontinuity in the F1 evolution is observable with a slight delay compared to F2, suggesting it is caused by a large scale disturbance propagating across the domain. The evolution of the transverse velocity component $v_{\mathrm{T}}$ shows several large amplitude disturbances at the apex of the loop with typical amplitudes of $20$ km s$^{-1}$; these are not mirrored by the velocity at the footpoints and therefore not driven by footpoint motions. Lack of any associated large amplitude deviations in the plasma density and temperature suggests that these deviations are caused by external disturbances, that is not originating in the loop. The fieldline F2 shows clear oscillation in the longitudinal velocity component $v_{\mathrm{L}}$ with $\sim$ 250 s period (marked as segment \textit{a} in Fig. \ref{fig:f2}). These periodic disturbances are propagating from the footpoint FP2 to footpoint FP1 at a speed of $\sim$ 90 km s$^{-1}$, that is close to the average local sound speed (we note that are also some signatures of propagation in opposite direction. The longitudinal oscillation is visible both in fieldline evolution plot and the wavelet spectra and lasts for most of the duration of the simulation sequence until $t = 2000$ s. The wavelet spectrum shows slight gradual decrease in period, probably linked to decrease in the total loop length. No clear attenuation is observable for the duration of the oscillation. The loop is also subject to shorter period transverse oscillations visible in the $v_{\mathrm{T}}$ velocity component with period of 180 s (segment \textit{c}), and similar oscillation can be also seen in the $v_{\mathrm{V}}$ evolution (segment \textit{d}). The oscillation in $v_{\mathrm{T}}$ follows a large initial displacement that occurred at $t = 900$ s and is rapidly attenuated in a similar manner to the large amplitude damped oscillations excited by impulsive events observed in the solar corona. In total 4 oscillation periods can be observed before the oscillation completely decays. Finally, we also note that high-frequency transverse oscillations can be observed in the loop during the initial 400 s with period less than 100 s (segment \textit{b}). The period changes during the oscillation duration, likely due to the change of the conditions in the loop which are linked to the condensation formation discussed earlier. In total, 5 oscillation periods are observable with no clear damping.

\subsection{Fieldline F3}
Fieldline F3 undergoes rapid expansion during the impulsive event at occurring at $t = 500$ s discussed above, during which the total length of the loop doubles (Fig. \ref{fig:f3}). This disturbance is also very clear in the evolution of the velocity component $v_{\mathrm{T}}$ at both footpoints and the loop apex and manifests in a transverse displacement with the velocity amplitude of 47 km s$^{-1}$ at the apex. An oscillation is visible in the transverse velocity component $v_{\mathrm{T}}$ (segment \textit{b} in Fig. \ref{fig:f3}) observable during the whole duration of the simulation with periodicity of $\sim$ 180 s, which changes slightly over time. This oscillation is also picked up in the evolution of the $v_{\mathrm{V}}$ component (segment \textit{c}), due to the nature of our coordinate system and the fact that transverse coronal loop oscillations can have any polarisation in the plane perpendicular to the magnetic field vector. There is no observable damping even following the impulsive event, and the amplitude of the transverse oscillation remains around $5$ km s$^{-1}$ during the whole duration of the simulation. The transverse oscillation is clearly present even before the impulsive event; this suggests that the impulsive event does not act as the primary driver of the oscillation. No persistent phase shift in single direction is observable between the $v_{\mathrm{T}}$ evolution at the loop apex and the two footpoints. The position of maximum  $v_{\mathrm{T}}$ amplitude varies between $s = 5$ Mm and $s = 10$ Mm along the loop. This suggests the oscillation corresponds to a standing transverse oscillation, as the velocity amplitude is greater at the loop apex than it is at close to the loop footpoints and there is no indication that the oscillation is driven by the footpoint motion. The observed oscillation mode likely corresponds to the fundamental harmonic, as only one oscillation antinode is observed. We note that despite the oscillatory behaviour being obvious from the plots of $v_{\mathrm{T}}$ and $v_{\mathrm{V}}$ evolution, it is not picked up by wavelet spectra above 95\% confidence level. Oscillations with similar period are present in the evolution of $v_{\mathrm{L}}$ velocity component (segment \textit{a}). Oscillatory behaviour of F3 is also observable in the time-distance plot using synthetic SDO/AIA 171 {\AA} observations shown in Fig. \ref{fig:fomo}. The position of the loop F3 in the time-distance plot starts at $y_{\mathrm{t}} = 5$ Mm along the slit and, following the expansion, gradually moves outwards and away from the centre of the domain towards $y_{\mathrm{t}} = 2.5$ Mm. From the forward-modelled intensity it is also clear that several loops in the vicinity of F3 are also subject to oscillations with similar period and duration as the F3 oscillation. These fieldlines are therefore likely part of a larger loop bundle that is subject to a collective transverse oscillation. 

\subsection{Fieldline F4}
Fieldline F4 initially corresponds to a short coronal loop with the initial length of 10 Mm (Fig. \ref{fig:f4}). During the first 400 s the loop expands nearly doubling its length. Another dramatic expansion of the loop starts after $t = 1500$ s, after which the loop apex eventually reaches the upper boundary and the loop becomes an open fieldline at $t = 1670$ s. An oscillation is observable in the longitudinal component of velocity $v_{\mathrm{L}}$ between $t = 350$ s and $t = 850$ s (segment \textit{a} in Fig. \ref{fig:f4}). The oscillation profile has a node close to the apex of the loop, reminiscent of second longitudinal harmonic. The period of the longitudinal oscillation is $\sim200$ s, and it increases over time. No clear periodic behaviour is visible in the evolution of $v_{\mathrm{T}}$ and $v_{\mathrm{V}}$ velocity components during the lifetime of the closed fieldline. After the fieldine opens, the total length integrated along the fieldline jumps to approximately half of its original value. Quasi-periodic disturbances propagating from the upper boundary downwards can be observed in the evolution of $v_{\mathrm{L}}$. These appear coupled to high frequency oscillatory behaviour with $\sim 60$ s period observable in $v_{\mathrm{T}}$ and $v_{\mathrm{V}}$ components for the remained of the lifetime of the open fieldline. As these are likely to be an artifact from the open boundary, we do not analyse them further. We also note that large amplitude transverse oscillation pattern is observable in the synthetic EUV emission close to the projected position of F4. as this does not seem to match with the evolution of the transverse velocity $v_{\mathrm{T}}$ of F4, we conclude that most of the emission comes from the lower lying loops. Again, several strands in this region can be seen to oscillate in phase, pointing to a collective transverse oscillation.

\subsection{Fieldline F5}
The evolution of fieldline F5 is shown in Fig. \ref{fig:f5}. This fieldline approaches F2 at $t = 1500$ s (Fig. \ref{fig:context_loops}), but for most of the duration of the simulation, the two loops evolve independently from each other without signs of collective behaviour. Similarly to fieldlines discussed above, F5 is also affected by the impulsive event at $t =  500$ s that sweeps across the coronal loops in the centre of the simulation domain ( Fig. \ref{fig:fomo}). Oscillatory behaviour in the longitudinal velocity component $v_{\mathrm{L}}$ with the spatial structure reminiscent of second harmonic is identifiable from $t = 500$ s to approximately $t =  1250$ s (segment \textit{a} in Fig. \ref{fig:f5}), after which the periodic evolution becomes less clear and harder to identify. The oscillation node lies slightly off-centre closer to FP1, this oscillatory behaviour is therefore also picked up in the wavelet spectrum for the evolution of $v_{\mathrm{L}}$ at the apex, although with evolving period. The wavelet spectra further show a strong signal for the periodicity in the vertical component of the velocity $v_{\mathrm{V}}$. Vertical oscillations with $\sim250$ s period are also clear in the fieldline evolution plot (segment \textit{c}). Oscillation with similar characteristics is also visible in the $v_{\mathrm{T}}$ component (segment \textit{b}), although less clearly. Both  $v_{\mathrm{T}}$ and  $v_{\mathrm{V}}$ oscillations have the points of maximum amplitude located in the upper parts of the loop, suggesting the oscillation is not driven by transverse displacement of the footpoints. This is further supported by the line plot of the evolution of the transverse velocity component, which shows that the velocity amplitude at the apex always dominates over velocity amplitude at the loop footpoints. We note that the evolution of the loop length is coupled to the oscillatory behaviour, as clearly visible from the evolution of the position of the footpoint FP2 in the loop evolution plot. This suggests that loop is undergoing a standing oscillation with with $\sim250$ s period polarised in vertical direction. Unlike horizontally polarised standing oscillations, these affect the total length of the loop and lead to periodic inflows/outflows from the loop centre as the loop expands/contracts. The oscillation cannot be indisputably linked to a single impulsive event, but it follows two successive temperature increases in the loop plasma occurring at $t = 300$ s and $t = 600$ s respectively.

\subsection{Footpoint evolution}
\begin{figure*}
	\includegraphics[width=43pc]{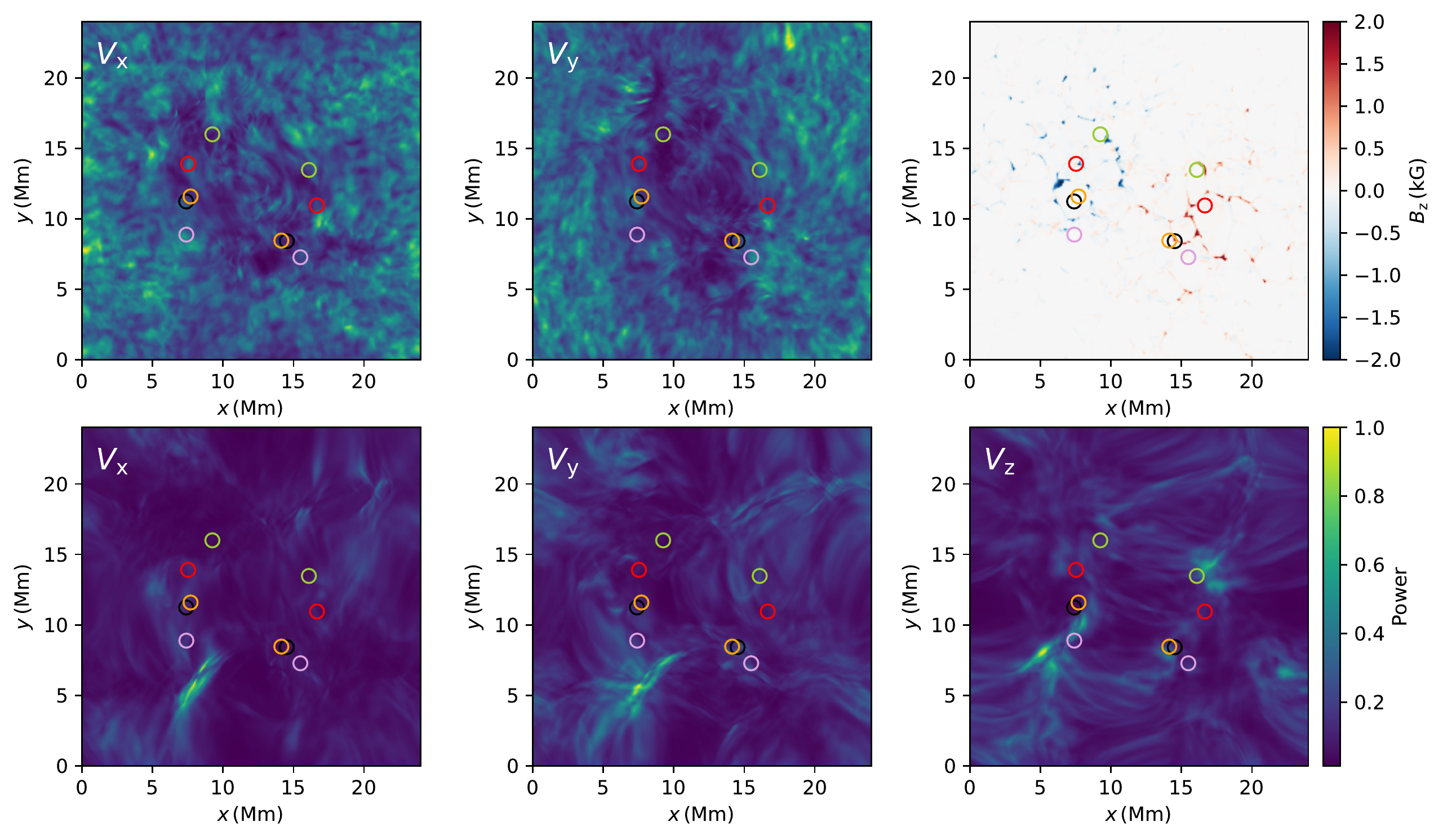}
	\caption{Top: Integrated power between 1.67 mHz and 16.7 mHz for $v_{\mathrm{x}}$ (left), $v_{\mathrm{y}}$ (centre) components of the velocity at $z = 1.4$ Mm in the chromophere and line-of-sight photospheric magnetic field at t = 180 s (right). Bottom: Integrated power between 1.67 mHz and 16.7 mHz for $v_{\mathrm{x}}$ (left), $v_{\mathrm{y}}$ (centre) and $v_{\mathrm{z}}$ (right) components of the velocity at $z = 10$ Mm in the corona. Coloured circles with 1 Mm diameter are centred on positions of footpoint coordinates at $z = 1.4$ Mm at $t = 0$ s and colour-coded as follows: F1-black, F2-red, F3-purple, F4-green, F5-orange. Animation of this figure for the full duration of the simulation is available.}
	\label{fig:pmaps}
\end{figure*}

We further calculate Fourier power spectra for each cell in the horizontal plane of the simulation box at two heights: at z = 1.4 Mm in the chromosphere and at z = 10 Mm in the corona. We sum the power for all frequencies between 1.67 mHz and 16.7 mHz, corresponding to 600 s and 60 s periods respectively, which covers the range of periods of oscillatory behaviour detected in the simulation. The resulting integrated power maps therefore highlight locations with increased oscillatory power regardless of the period of the oscillation; we compare these with the instantaneous positions of magnetic loop footpoints (Fig. \ref{fig:pmaps}).

This evolution highlights the dynamic nature of coronal loops, as their footpoints are not static but evolve during the entire simulation as they get dragged around by convective flows. Such changes occur on the same timescales as the evolution of line-of-sight component of the photospheric magnetic field. Occasionally the evolution is more rapid, most likely when involving impulsive reconnection events at coronal heights. In general, footpoints of oscillating loops are not embedded in locations with increased oscillatory power. Harmonic driver is therefore not a prerequisite for onset of coronal loop oscillations.

\section{Discussion}
An important point for consideration is the applicability of the fieldline tracing method that is used to study the fieldline evolution in the instances of magnetic reconnection. Tracing of the magnetic fieldlines is equivalent to tracing the grid points in the simulation domain which are magnetically connected. Since in the corona the matter and energy transport happens along the magnetic field, tracing the magnetic fieldlines is therefore the closest we can get to studying the true evolution of coronal structures without being affected by line of sight effects, which can significantly influence determination of oscillation parameters \citep[see e.g.][]{moortel_2012}. When a magnetic loop ‘reconnects’, the magnetic connectivity of a certain part of the loop changes. However, advecting the seed points initially placed into regions where oscillations were detected ensures we keep tracing the magnetic connectivity of the oscillating part even if reconnection occurs in other parts of the loop. We therefore argue that this approach remains the most feasible way of tracing the true evolution of coronal structures.

We note that evolution in line-of-sight integrated emission does not necessarily copy the evolution of traced magnetic fieldlines with the same $x,y$ coordinates. This is likely due to line-of-sight superposition of emission from several magnetic structures, which can vary in emissivity and in presence or lack of oscillatory behaviour. This effect of line-of-sight superposition on oscillation analysis has been discussed by previous studies \citep{moortel_2012, leenaarts_2015}. Our work further highlights the limitations posed by line-of-sight effects which should be taken into account when analysing observations of coronal oscillations.

The identification of observed transverse oscillations as either standing and propagating is complicated by the relatively short length of coronal loops in the simulation and by high typical values of kink speed in the corona which translates to high phase speed of the oscillation. Assuming a mean kink speed in the range 100 - 1000 km s$^{-1}$, the expected propagation time along a 20 Mm long loop varies from 200 s to 20 s, corresponding to 20 and 2 time steps of the simulation output, respectively. The position of the maximum oscillation amplitude higher up along the loop rather than at the loop footpoints suggests the oscillations are in fact standing with the loop footpoints acting as oscillation nodes. The term nodes is however used loosely here, since the footpoints are in fact not static, but continuously moving (Fig. \ref{fig:pmaps}). There are no velocity nodes clearly observable in the coronal sections of studied loops and the longitudinal profile of $V_{\mathrm{T}}$ component matches fundamental oscillation harmonic. We note that similar oscillations in chromospheric fibrils analysed in synthetised H-alpha emission have been identified as propagating by \citet{leenaarts_2015}. The method commonly used to investigate presence of oscillation phase shift and the corresponding phase speed is based on detecting phase shift in time-distance plots taken at several different positions along the studied loop, using observed or forward-modelled emission. It should be noted however, that such method only works for static loops. It is likely to produce spurious results in the presence of the bulk motion of the fieldline across of the slits in addition to the oscillatory motion, which seems to be the case for all 5 fieldines studied in this work. Instead we focused of the evolution of the oscillation phase at different locations along each loop that have been traced with time, while accounting for the loop 3D structure and evolution. No persistent phase shift has been identified between the loop apex and the footpoints in any of the analysed cases, that would suggest observed transverse oscillations are propagating.

On the other hand, such distinction is easily made for longitudinal oscillations as the typical values of the sound speed in the corona are order of magnitude smaller compared to Alfv\'{e}n  or kink speed (see Table \ref{tab:speeds}). Longitudinal oscillations propagating from one footpoint to another with local sound speed were detected in F2. The mean time delay between the three observable successive wave trains is around 250 s. Such propagating waves can be a result of an acoustic driver (potentially linked to the global p-mode equivalent oscillations) which drives compressive longitudinal waves that steepen into shocks in the chromosphere and propagate through the corona. Such propagating oscillations were however not universally present for all of the studied fieldlines, hence we refrain from drawing any conclusions about the periodicity (or lack thereof) of the driver. We note that the period of the global p-mode oscillation seen in the simulation is greater, around 450 s, as discussed in section \ref{section:model}. Standing longitudinal oscillations sustained for few oscillation periods were detected in F1, F3, F4 and F5, with relatively short periods between 100-200 s, depending on the loop. The longitudinal velocity profiles showing an oscillation node at the apex of the loops suggest these correspond to a second harmonic and likely represent a response of the loop to external perturbations occurring at coronal heights \citep{nakariakov_2004}. In the loops that show the presence of second harmonic of a longitudinal standing mode the onset of the oscillation follows events associated with increase of temperature of the plasma, and hence likely linked to small scale reconnection events/other impulsive heating events. Oscillation damping is observable for most cases, however, detailed analysis of the damping profiles and associated dissipation is beyond the scope of this work.

\begin{figure}
	\includegraphics[width=21pc]{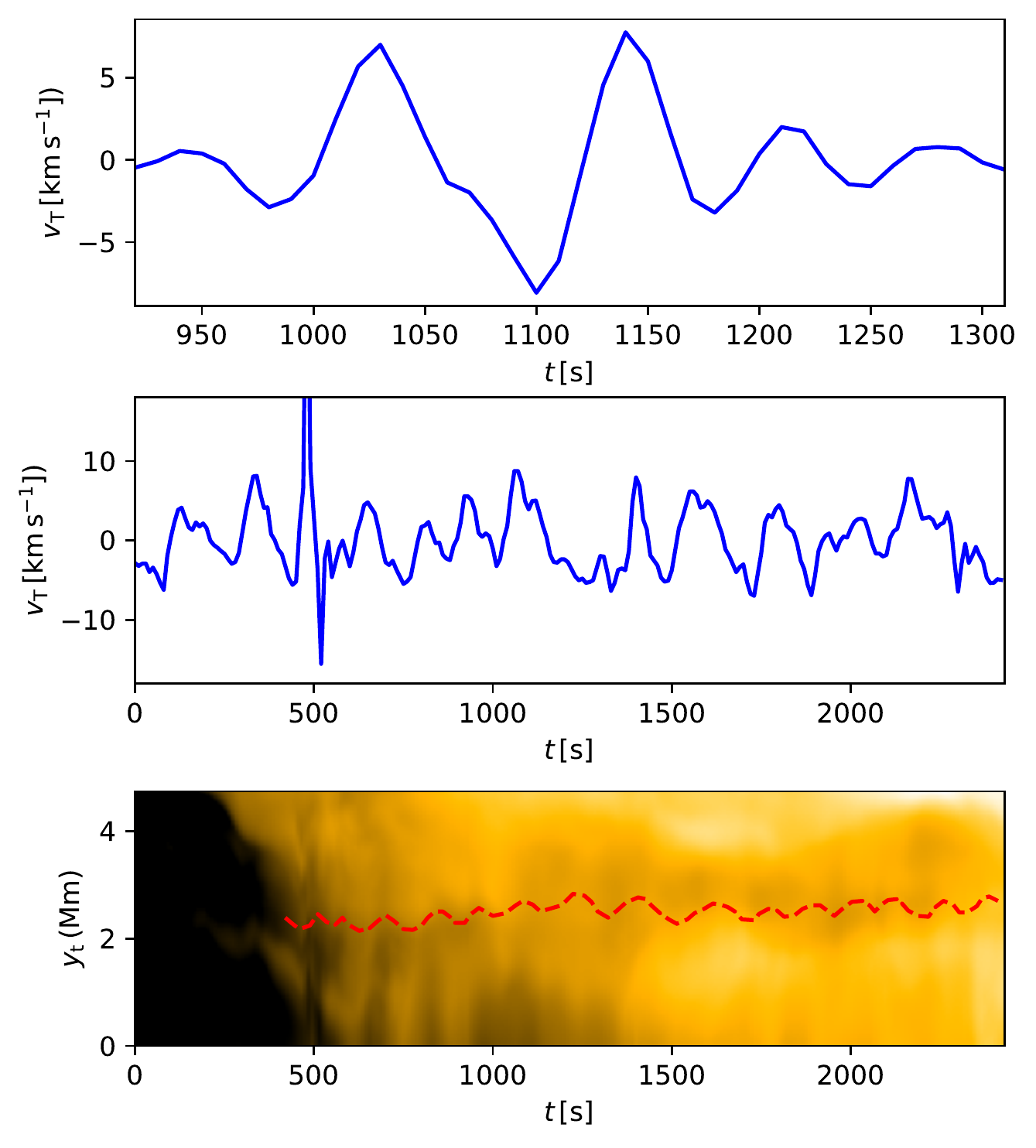}
	\caption{Top: Evolution of $v_{\mathrm{T}}$ component (blue) at the apex of loop F2 showing a damped oscillation following an event possibly associated with impulsive temperature increase. A large-scale trend was removed from $v_{\mathrm{T}}$ time series by fitting and subtracting a second degree polynomial. Middle: Evolution of $v_{\mathrm{T}}$ component at the apex of loop F3 showing a sustained oscillation. Bottom: Sustained oscillation in the time-distance plot of forward-modelled  SDO/AIA 171 {\AA} emission.}
	\label{fig:regimes}
\end{figure}

We also highlight the evolution of $v_{\mathrm{T}}$ in F3, which shows clear oscillation with 180 s period that is sustained over time and does not show any observable damping (Fig. \ref{fig:regimes}). There is no consistent phase shift between the oscillatory behaviour at the loop apex and in the loop legs, suggesting the oscillation is standing. This is further supported by the fact that the location of the maximum velocity amplitude lies close to the loop apex. This oscillation pattern is very similar to the commonly observed regime of decayless oscillations \citep{wang_2012, nistico_2013, anfinogentov_2015}.
Conversely, an example of the damped impulsively generated (or "decaying") oscillation regime is observable in F2 following an event associated with impulsive temperature increase in the loop (Fig. \ref{fig:regimes}). The oscillation velocity amplitude is largest at the apex and the evolution of the longitudinal $V_{\mathrm{T}}$ profile matches fundamental harmonic of a standing transverse oscillation. This suggests it is a natural response of the loop to a perturbation accompanied by an impulsive energy release. We also note that for several cases of oscillations in loops F1, F4 and F5, the classification into "decaying" and "decayless" is not as clear (due to lack of clear steady damping pattern such as shown in Fig. \ref{fig:regimes}). F3 is the only loop that shows persistent undamped oscillation present for the whole duration of simulation.

The oscillatory motions seen in the simulation in magnetic structures, both "decaying" and "decayless" regimes correspond to standing transverse oscillations with the oscillation antinodes lying at coronal heights and are therefore consistent with the observations of transverse oscillations of coronal loops as seen in TRACE and SDO/AIA data \citep[e.g.][]{aschwanden_1999, white_2012}, where the standing transverse oscillation are the most commonly observed regime.

We further note that the detected oscillation periods vary between different fieldlines and can also change with time, particularly in loops that are subject to change in physical properties, or with increasing/decreasing loop length, as would be expected for standing oscillations (this is true for all oscillation modes visible in the simulation). This spread and variability of the oscillation periods as well as lack of oscillation coherence in different parts of the simulation domain suggests that they are not a result of a coherent global oscillation in the simulation. Our analysis therefore does not agree with the premise that the transverse oscillations in the corona are driven by the global p-mode oscillation which is based on comparison of peaks in the power spectra from velocity measurements in the corona and the p-mode power spectra \citep{morton_2016, morton_2019} (it should however be noted that analysis in such studies focuses on very long coronal loops, i.e. the spatial scales are very different from the short low-lying loops studied in this work). Such mechanism requires coronal loops to oscillate with the same period regardless of their length, as they are driven at their footpoints by a harmonic driver. The variability of the oscillation period in the simulation therefore excludes the presence of a harmonic driver, or at the very least suggests that such driver is not a dominant mechanism responsible for the excitation of transverse oscillations in this work. We stress that care should be taken when drawing conclusions from the global power spectra due to the temporal variability of the oscillation periods of individual magnetic structures which has been shown both observationally \citep{verwichte_2017} and by numerical simulations \citep{kohutova_2017}.

Excitation mechanism proposed by \citet{nakariakov_2016} is based on identifying the decayless oscillations as self-oscillations of the coronal loop, that are excited by interaction with quasi-steady flows, similar to a bow moving across a violin string. The flows in the simulation domain are too dynamic and not stable enough to drive persistent self-oscillation of the fieldlines; this is evidenced by the motion of loop footpoints that are dragged around by turbulent flows in the lower solar atmosphere shown in Fig. \ref{fig:pmaps} and show variability on the order of minutes. 3D numerical models however seem to suggest that excitation of self-oscillations requires persistent flows that are steady on timescales of the order of an hour \citep{karampelas_2020}.   

The absence of increased oscillatory power in the chromosphere at the positions of the footpoints of the oscillating fieldlines also suggests that a harmonic footpoint driver is not a prerequisite for the excitation of coronal oscillations. Open questions addressing the excitation of the decayless oscillations therefore still remain. If their excitation mechanism is global, why are they not observed in all coronal loops? The decayless oscillations are sufficiently abundant to not be driven by an isolated event \citep{anfinogentov_2015}, but not enough to classify global process as the driver.  

Finally, the differences between the magnetic and density structure of the simulated and real solar corona, and hence the limitations of the model used for our analysis need to be addressed. The forward modelled EUV emission from the simulated corona is comparatively more diffuse and lacks a lot of fine-scale structure seen in real coronal observations, especially fine-scale coronal strands revealed by observations from the second Hi-C flight \citep{williams_2020}, which are not resolved in SDO/AIA observations. The simulated corona also suffers from a lack of coronal loops understood in the traditional sense as distinct structures denser than the surroundings with distinct boundaries. As the simulation does not include any flux emergence, the coronal loops in the simulation do not correspond to overdense flux tubes lifted into the corona from the lower solar atmosphere but are instead filled with evaporated plasma due to impulsive heating events. Due to enhanced heating regions having an irregular shape \citep{kohutova_2020b} the structures filled with evaporated plasma have greater spatial extent and lack well-defined boundaries. Simulation resolution might also be a limiting factor, eventhough the characteristic transverse scales seen in time-distance plots are well above the horizontal grid size of 48 km. Distinct oscillating strands are however still observable in Fig. \ref{fig:fomo}. Furthermore, recent numerical studies of evolution of initially homogeneous coronal loops as a response to transverse motions suggest that our highly idealised picture of coronal loops as monolithic plasma cylinders is not very realistic in the first place \citep{magyar_2018, karampelas_2019, antolin_2019}.

\section{Conclusions}

We have studied the excitation and evolution of coronal oscillations in 3D self-consistent simulations of solar atmosphere spanning from convection zone to solar corona using radiation-MHD code Bifrost. We have combined forward-modelled EUV emission with 3D-tracing of magnetic field though both space and time in order to analyse oscillatory behaviour of individual magnetic loops, while accounting for their dynamic evolution. We have analysed evolution of different velocity components, using wavelet analysis to capture changes in the oscillatory periods due to evolution of properties of the oscillating magnetic loops. Various types of oscillations commonly observed in the corona are reproduced in such simulations. We detected standing oscillations in both transverse and longitudinal velocity components, including higher order oscillation harmonics. We have further shown that self-consistent simulations reproduce existence of two distinct regimes of transverse coronal oscillations, that is rapidly decaying oscillations triggered by impulsive events and sustained small-scale oscillations showing no observable damping. Damped transverse oscillations were found to be associated with instances of impulsive energy release, such as small scale reconnection events, or external perturbations leading to sudden changes in the loop properties, in agreement with coronal observations. Persistent transverse oscillations on the other hand were not linked to any such impulsive events. We did not find any evidence for this oscillation regime being driven by a global (simulation) p-mode. Lack of enhanced oscillatory power near the footpoint regions of the studied loops together with variability of oscillation periods between different coronal loops and lack of oscillation coherence across the simulation domain exclude any type of harmonic driver as being responsible for excitation of the oscillations. 

Our work therefore highlights the complexity of coronal oscillations in simulations with realistic magnetic field configurations which include the complex dynamics of the lower solar atmosphere. Care needs to be taken when translating findings from highly idealised models into real solar observations, as there are several limitations to treating coronal loops as static structures. We have shown that individual fieldlines are very dynamic, their footpoints migrate and their overall length changes significantly over realistic timescales. This might have non-negligible consequences for accuracy of estimates of coronal plasma parameters deduced using coronal seismology. The oscillating coronal structures we analysed are far from idealised plasma cylinders. The focus of modelling work should therefore be shifting from rigid cylindrical models towards more realistic descriptions that account for rapid variability, complex morphology and presence of nonlinearities. There are obviously natural limitations to this approach resulting from the computational expense of building such models and also from the associated limits on physical size of the simulation domain. Work is currently underway to investigate the evolution of coronal loops in 3D self-consistent simulations that span high into the corona. We would hence like to highlight the potential of using self-consistent simulations of the solar atmosphere as a laboratory for testing assumptions made by coronal seismology and models of various damping and dissipation mechanisms in an environment with realistic density structure and magnetic field geometry.

\begin{acknowledgements}
This research was supported by the Research Council of Norway through its Centres of Excellence scheme, project no. 262622. 
\end{acknowledgements} 

\bibliography{waves}

\begin{thebibliography}{57}
\expandafter\ifx\csname natexlab\endcsname\relax\def\natexlab#1{#1}\fi

\bibitem[{Afanasyev {et~al.}(2020)Afanasyev, Van~Doorsselaere, \&
  Nakariakov}]{afanasyev_2020}
Afanasyev, A.~N., Van~Doorsselaere, T., \& Nakariakov, V.~M. 2020, A\&A, 633,
  L8

\bibitem[{Anfinogentov {et~al.}(2015)Anfinogentov, Nakariakov, \&
  Nisticò}]{anfinogentov_2015}
Anfinogentov, S.~A., Nakariakov, V.~M., \& Nisticò, G. 2015, A\&A, 583, A136

\bibitem[{Antolin \& Van~Doorsselaere(2019)}]{antolin_2019}
Antolin, P. \& Van~Doorsselaere, T. 2019, Front. Phys., 7

\bibitem[{Antolin {et~al.}(2014)Antolin, Yokoyama, \&
  Van~Doorsselaere}]{antolin_2014}
Antolin, P., Yokoyama, T., \& Van~Doorsselaere, T. 2014, ApJL, 787, L22

\bibitem[{Aschwanden {et~al.}(1999)Aschwanden, Fletcher, Schrijver, \&
  Alexander}]{aschwanden_1999}
Aschwanden, M.~J., Fletcher, L., Schrijver, C.~J., \& Alexander, D. 1999, ApJ,
  520, 880

\bibitem[{Berghmans \& Clette(1999)}]{berghmans_1999}
Berghmans, D. \& Clette, F. 1999, Sol. Phys., 186, 207

\bibitem[{Carlsson {et~al.}(2010)Carlsson, Hansteen, \&
  Gudiksen}]{carlsson_2010}
Carlsson, M., Hansteen, V.~H., \& Gudiksen, B.~V. 2010, Mem. Soc. Ast. It., 81,
  582

\bibitem[{Carlsson {et~al.}(2016)Carlsson, Hansteen, Gudiksen, Leenaarts, \&
  De~Pontieu}]{carlsson_2016}
Carlsson, M., Hansteen, V.~H., Gudiksen, B.~V., Leenaarts, J., \& De~Pontieu,
  B. 2016, A\&A, 585, A4

\bibitem[{Chen \& Peter(2015)}]{chen_2015}
Chen, F. \& Peter, H. 2015, A\&A, 581, A137

\bibitem[{De~Moortel(2005)}]{moortel_2005}
De~Moortel, I. 2005, Phil. Trans. R. Soc. A, 363, 2743

\bibitem[{De~Moortel {et~al.}(2000)De~Moortel, Ireland, \&
  Walsh}]{moortel_2000}
De~Moortel, I., Ireland, J., \& Walsh, R.~W. 2000, A\&A, 355, L23

\bibitem[{De~Moortel \& Nakariakov(2012)}]{de_moortel_2012}
De~Moortel, I. \& Nakariakov, V.~M. 2012, Phil. Trans. R. Soc. A, 370, 3193

\bibitem[{De~Moortel \& Pascoe(2012)}]{moortel_2012}
De~Moortel, I. \& Pascoe, D.~J. 2012, ApJ, 746, 31

\bibitem[{De~Pontieu {et~al.}(2004)De~Pontieu, Erdélyi, \&
  James}]{pontieu_2004}
De~Pontieu, B., Erdélyi, R., \& James, S.~P. 2004, Nature, 430, 536

\bibitem[{Duckenfield {et~al.}(2018)Duckenfield, Anfinogentov, Pascoe, \&
  Nakariakov}]{duckenfield_2018}
Duckenfield, T., Anfinogentov, S.~A., Pascoe, D.~J., \& Nakariakov, V.~M. 2018,
  ApJL, 854, L5

\bibitem[{Edwin \& Roberts(1983)}]{edwin_1983}
Edwin, P.~M. \& Roberts, B. 1983, Sol. Phys., 88, 179

\bibitem[{Eklund {et~al.}(2020)Eklund, Wedemeyer, Snow, Jess, Jafarzadeh,
  Grant, Carlsson, \& Szydlarski}]{eklund_2020}
Eklund, H., Wedemeyer, S., Snow, B., {et~al.} 2020, arXiv:2008.05324
  [astro-ph], arXiv: 2008.05324

\bibitem[{Felipe(2012)}]{felipe_2012}
Felipe, T. 2012, ApJ, 758, 96

\bibitem[{González-Morales {et~al.}(2019)González-Morales, Khomenko, \&
  Cally}]{gonzalez_2019}
González-Morales, P.~A., Khomenko, E., \& Cally, P.~S. 2019, ApJ, 870, 94

\bibitem[{Goossens {et~al.}(2019)Goossens, Arregui, \&
  Van~Doorsselaere}]{goossens_2019}
Goossens, M.~L., Arregui, I., \& Van~Doorsselaere, T. 2019, Front. Astron.
  Space Sci., 6

\bibitem[{Gudiksen {et~al.}(2011)Gudiksen, Carlsson, Hansteen, Hayek,
  Leenaarts, \& Martínez-Sykora}]{gudiksen_2011}
Gudiksen, B.~V., Carlsson, M., Hansteen, V.~H., {et~al.} 2011, A\&A, 531, A154

\bibitem[{Karampelas \& Van~Doorsselaere(2020)}]{karampelas_2020}
Karampelas, K. \& Van~Doorsselaere, T. 2020, ApJL, 897, L35

\bibitem[{Karampelas {et~al.}(2017)Karampelas, Van~Doorsselaere, \&
  Antolin}]{karampelas_2017}
Karampelas, K., Van~Doorsselaere, T., \& Antolin, P. 2017, A\&A, 604, A130

\bibitem[{Karampelas {et~al.}(2019)Karampelas, Van~Doorsselaere, \&
  Guo}]{karampelas_2019}
Karampelas, K., Van~Doorsselaere, T., \& Guo, M. 2019, A\&A, 623, A53

\bibitem[{Kohutova {et~al.}(2020{\natexlab{a}})Kohutova, Antolin, Popovas,
  Szydlarski, \& Hansteen}]{kohutova_2020b}
Kohutova, P., Antolin, P., Popovas, A., Szydlarski, M., \& Hansteen, V.~H.
  2020{\natexlab{a}}, A\&A, 639, A20

\bibitem[{Kohutova \& Verwichte(2017)}]{kohutova_2017}
Kohutova, P. \& Verwichte, E. 2017, A\&A, 606, A120

\bibitem[{Kohutova \& Verwichte(2018)}]{kohutova_2018}
Kohutova, P. \& Verwichte, E. 2018, A\&A, 613, L3

\bibitem[{Kohutova {et~al.}(2020{\natexlab{b}})Kohutova, Verwichte, \&
  Froment}]{kohutova_2020a}
Kohutova, P., Verwichte, E., \& Froment, C. 2020{\natexlab{b}}, A\&A, 633, L6

\bibitem[{Leenaarts {et~al.}(2015)Leenaarts, Carlsson, \& Rouppe van~der
  Voort}]{leenaarts_2015}
Leenaarts, J., Carlsson, M., \& Rouppe van~der Voort, L. 2015, ApJ, 802, 136

\bibitem[{Liu {et~al.}(2019)Liu, Nelson, Snow, Wang, \& Erdélyi}]{liu_2019}
Liu, J., Nelson, C.~J., Snow, B., Wang, Y., \& Erdélyi, R. 2019, Nat. Commun.,
  10, 3504

\bibitem[{Magyar \& Van~Doorsselaere(2016)}]{magyar_2016}
Magyar, N. \& Van~Doorsselaere, T. 2016, A\&A, 595, A81

\bibitem[{Magyar \& Van~Doorsselaere(2018)}]{magyar_2018}
Magyar, N. \& Van~Doorsselaere, T. 2018, ApJ, 856, 144

\bibitem[{Magyar {et~al.}(2015)Magyar, Van~Doorsselaere, \&
  Marcu}]{magyar_2015}
Magyar, N., Van~Doorsselaere, T., \& Marcu, A. 2015, A\&A, 582, A117

\bibitem[{Morton {et~al.}(2016)Morton, Tomczyk, \& Pinto}]{morton_2016}
Morton, R.~J., Tomczyk, S., \& Pinto, R.~F. 2016, ApJ, 828, 89

\bibitem[{Morton {et~al.}(2019)Morton, Weberg, \& McLaughlin}]{morton_2019}
Morton, R.~J., Weberg, M.~J., \& McLaughlin, J.~A. 2019, Nat. Astron., 3, 223

\bibitem[{Nakariakov {et~al.}(2016)Nakariakov, Anfinogentov, Nisticò, \&
  Lee}]{nakariakov_2016}
Nakariakov, V.~M., Anfinogentov, S.~A., Nisticò, G., \& Lee, D.-H. 2016, A\&A,
  591, L5

\bibitem[{Nakariakov \& Kolotkov(2020)}]{nakariakov_2020}
Nakariakov, V.~M. \& Kolotkov, D.~Y. 2020, Annu. Rev. Astron. Astrophys., 58,
  441

\bibitem[{Nakariakov {et~al.}(1999)Nakariakov, Ofman, Deluca, Roberts, \&
  Davila}]{nakariakov_1999}
Nakariakov, V.~M., Ofman, L., Deluca, E.~E., Roberts, B., \& Davila, J.~M.
  1999, Science, 285, 862

\bibitem[{Nakariakov {et~al.}(2004)Nakariakov, Tsiklauri, Kelly, Arber, \&
  Aschwanden}]{nakariakov_2004}
Nakariakov, V.~M., Tsiklauri, D., Kelly, A., Arber, T.~D., \& Aschwanden, M.~J.
  2004, A\&A, 414, L25

\bibitem[{Nakariakov \& Verwichte(2005)}]{nakariakov_2005}
Nakariakov, V.~M. \& Verwichte, E. 2005, Liv. Rev. Sol. Phys., 2, 3

\bibitem[{Nisticò {et~al.}(2013)Nisticò, Nakariakov, \&
  Verwichte}]{nistico_2013}
Nisticò, G., Nakariakov, V.~M., \& Verwichte, E. 2013, A\&A, 552, A57

\bibitem[{Pagano \& De~Moortel(2017)}]{pagano_2017}
Pagano, P. \& De~Moortel, I. 2017, A\&A, 601, A107

\bibitem[{Pagano \& De~Moortel(2019)}]{pagano_2019}
Pagano, P. \& De~Moortel, I. 2019, A\&A, 623, A37

\bibitem[{Peter {et~al.}(2013)Peter, Bingert, Klimchuk, De~Forest, Cirtain,
  Golub, Winebarger, Kobayashi, \& Korreck}]{peter_2013}
Peter, H., Bingert, S., Klimchuk, J.~A., {et~al.} 2013, A\&A, 556, A104

\bibitem[{Riedl {et~al.}(2019)Riedl, Van~Doorsselaere, \&
  Santamaria}]{riedl_2019}
Riedl, J.~M., Van~Doorsselaere, T., \& Santamaria, I.~C. 2019, A\&A, 625, A144

\bibitem[{Santamaria {et~al.}(2015)Santamaria, Khomenko, \&
  Collados}]{santamaria_2015}
Santamaria, I.~C., Khomenko, E., \& Collados, M. 2015, A\&A, 577, A70

\bibitem[{Shelyag {et~al.}(2011)Shelyag, Keys, Mathioudakis, \&
  Keenan}]{shelyag_2011}
Shelyag, S., Keys, P., Mathioudakis, M., \& Keenan, F.~P. 2011, A\&A, 526, A5

\bibitem[{Stein \& Nordlund(2001)}]{stein_2001}
Stein, R.~F. \& Nordlund, {\AA}. 2001, ApJ, 546, 585

\bibitem[{Tomczyk \& McIntosh(2009)}]{tomczyk_2009}
Tomczyk, S. \& McIntosh, S.~W. 2009, ApJ, 697, 1384

\bibitem[{Torrence \& Compo(1998)}]{torrence_1998}
Torrence, C. \& Compo, G.~P. 1998, Bull. Amer. Meteorol. Soc., 79, 61

\bibitem[{Van~Doorsselaere {et~al.}(2016)Van~Doorsselaere, Antolin, Yuan,
  Reznikova, \& Magyar}]{doorsselaere_2016}
Van~Doorsselaere, T., Antolin, P., Yuan, D., Reznikova, V., \& Magyar, N. 2016,
  Front. Astron. Space Sci., 3

\bibitem[{Verwichte \& Kohutova(2017)}]{verwichte_2017}
Verwichte, E. \& Kohutova, P. 2017, A\&A, 601, L2

\bibitem[{Verwichte {et~al.}(2004)Verwichte, Nakariakov, Ofman, \&
  Deluca}]{verwichte_2004}
Verwichte, E., Nakariakov, V.~M., Ofman, L., \& Deluca, E.~E. 2004, Sol. Phys.,
  223, 77

\bibitem[{Verwichte {et~al.}(2013)Verwichte, Van~Doorsselaere, Foullon, \&
  White}]{verwichte_2013}
Verwichte, E., Van~Doorsselaere, T., Foullon, C., \& White, R.~S. 2013, ApJ,
  767, 16

\bibitem[{Wang {et~al.}(2012)Wang, Ofman, Davila, \& Su}]{wang_2012}
Wang, T., Ofman, L., Davila, J.~M., \& Su, Y. 2012, ApJ, 751, L27

\bibitem[{White \& Verwichte(2012)}]{white_2012}
White, R.~S. \& Verwichte, E. 2012, A\&A, 537, A49

\bibitem[{Williams {et~al.}(2020)Williams, Walsh, Winebarger, Brooks, Cirtain,
  De~Pontieu, Golub, Kobayashi, McKenzie, Morton, Peter, Rachmeler, Savage,
  Testa, Tiwari, Warren, \& Watkinson}]{williams_2020}
Williams, T., Walsh, R.~W., Winebarger, A.~R., {et~al.} 2020, ApJ, 892, 134

\end{thebibliography}

\end{document}